\documentstyle[a4,epsfig,12pt]{article}

\begin{document}
\begin{center}

{\large {\bf Relativistic Nucleus-Nucleus Collisions: A Connection
between the Strangeness-Maximum at $\sqrt{s} \approx$ 7 GeV and
the QCD Critical Endpoint from Lattice Studies}}

\vspace{0.3cm}
 {\bf R. Stock, University of Frankfurt, Germany}
\end{center}

\begin{abstract}
 A steep maximum occurs in the Wroblewski ratio between strange
and non-strange quarks created in central nucleus-nucleus
collisions, of about A=200, at the lower SPS energy $\sqrt{s}
\approx$ 7 GeV. By analyzing hadronic multiplicities within the
grand canonical statistical hadronization model this maximum is
shown to occur at a baryochemical potential of about 450 MeV. In
comparison, recent QCD lattice calculations at finite
baryochemical potential suggest a steep maximum of the light quark
susceptibility, to occur at similar $\mu_B$, indicative of
"critical fluctuation" expected to occur at or near the QCD
critical endpoint. This endpoint hat not been firmly pinned down
but should occur in the 300 MeV $ < \mu_B^c<700$ MeV interval. It
is argued that central collisions within the low SPS energy range
should exhibit a turning point between compression/heating, and
expansion/cooling at energy density, temperature and $\mu_B$ close
to the suspected critical point. Whereas from top SPS to RHIC
energy the primordial dynamics create a turning point far above in
$\epsilon$ and T, and far below in $\mu_B$. And at lower AGS
energies the dynamical trajectory stays below the phase boundary.
Thus, the observed sharp strangeness maximum might coincide with
the critical $\sqrt{s}$ at which the dynamics settles at, or near
the QCD endpoint.
\end{abstract}

Bulk hadron production systematics in central nucleus-nucleus
collisions at relativistic energy is, overall, well reproduced by
a statistical Hagedorn hadronic freeze-out model. A grand
canonical version of this model captures the various hadronic
species multiplicities, per collision event, from pions to omega
hyperons, in terms of a few universal parameters that describe the
dynamical stage in which the emerging hadronic matter decays to a
quasi-classical gas of free resonances and hadrons \cite{1,2,3}.
The grand canonical parameters are temperature T, volume V and
chemical potential $\mu$. They capture a snapshot of the fireball
expansion within the narrow time interval surrounding hadronic
chemical freeze-out, which thus appears to populate the
hadron/resonance mass and quantum number spectrum, predominantly,
by phase space weight \cite{4,5} thus creating an apparent thermal
equilibrium state prevailing in the produced
hadron-resonance-population. This chemical equilibrium
instantaneously decouples from fireball expansion surviving
further (near isentropic) processes. It can thus be retrieved from
the finally observed hadronic multiplicities, by state of the art
grand canonical model analysis. This analysis succeeds from AGS,
via SPS, to RHIC energy \cite{1,2,3,5}.

Statistical model analysis is also applicable to elementary
collisions, $p+p$, $p+\overline{p}$, and $e^+e^-$ annihilation as
was shown by Hagedorn \cite{6} and, more recently, by Becattini
and Collaborators \cite{7,8}. The canonical version of ensemble
analysis is applicable here. Mutatis mutandis the same
hadrochemical equilibrium feature is being attested, emphasizing
the statement that the apparent equilibrium does not arise from a
thermodynamical inelastic rescattering cascade toward equilibrium
- there is essentially none in elementary processes - but should
stem directly from the QCD hadronization process occuring under
phase space dominance \cite{4,5}.

The crucial difference between elementary and central
nucleus-nucleus resides, in statistical model view, in a
transition from canonical to grand canonical order in the ensuing
decoupled hadronic state. This transition was studied by Cleymans,
Tounsi, Redlich et al. \cite{9}. Its main feature is strangeness
enhancement.Comparing the strange to non-strange hadron
multiplicities in elementary, and in central nucleus-nucleus
collisions at similar energy, one observes an increase of the
singly strange hyperons and mesons, relative to pions, of about
2-4, and corresponding higher relative enhancements of multiply
strange hyperons \cite{10,11,12}, ranging up to order-of-magnitude
enhancement. In the terminology of Hagedorn statistical models,
strangeness is suppressed in the small system, canonical case, of
elementary collisions (due to the dictate of local strangeness
conservation in a small "fireball" volume), whereas it approaches
flavour equipartition in large fireballs due to the occurence of
quantum number conservation, on average only, over a large volume
- as reflected by the {\bf global} chemical potential featured by
the grand canonical ensemble: "strangeness enhancement" occurs as
the fading-away of canonical constraints.

From statistical model analysis we obtain a more general view of
strangeness relative to non-strangeness production than is
provided by considering individual strange to non-strange
production ratios, like $K/\pi, \:\Omega/\pi$ etc., from $p+p$ to
central A+A. The model quantifies strange to non-strange
hadron/resonance production by means of Wroblewski quark counting
at hadronic freeze-out \cite{13}. It determines the so-called
Wroblewski-ratio,
\begin{equation}
\lambda_s=\frac{2(<s>+<\overline{s}>)}{<u>+<\overline{u}>+<d>+<\overline{d}>}
\end{equation}

which quantifies the overall strangeness to non-strangeness ratio
at hadronic freeze-out. Strangeness enhancement (i.e. removal of
strangeness suppression in elementary collisions) is quantified,
by such an analysis, to proceed from $\lambda_s \approx$0.25 in
elementary collisions, to $\lambda \approx$ 0.45 in central
nucleus-nucleus collisions \cite{2,3}.

\begin{figure}
\begin{center}
\epsfig{figure=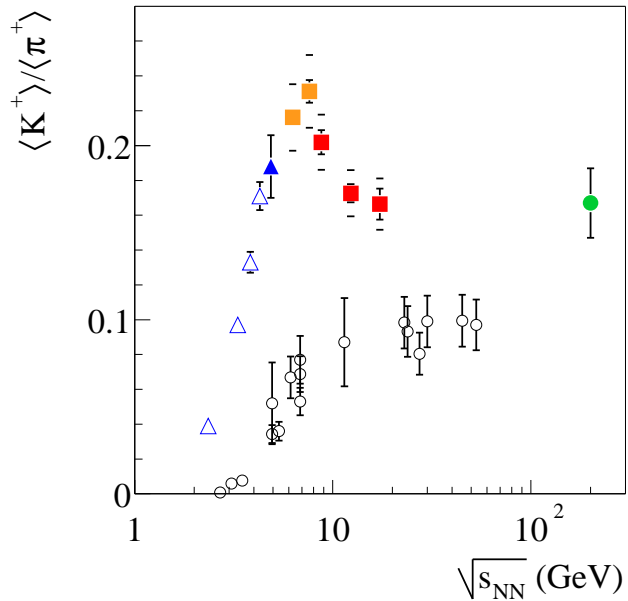,width=90mm}
\end{center}
\caption{Energy dependence of $<K^+>/<\pi^+>$ ratio for central
Pb+Pb (Au+Au) collisions (upper points) and p+p interactions
(lower points).}
\end{figure}

\begin{figure}
\begin{center}
\epsfig{figure=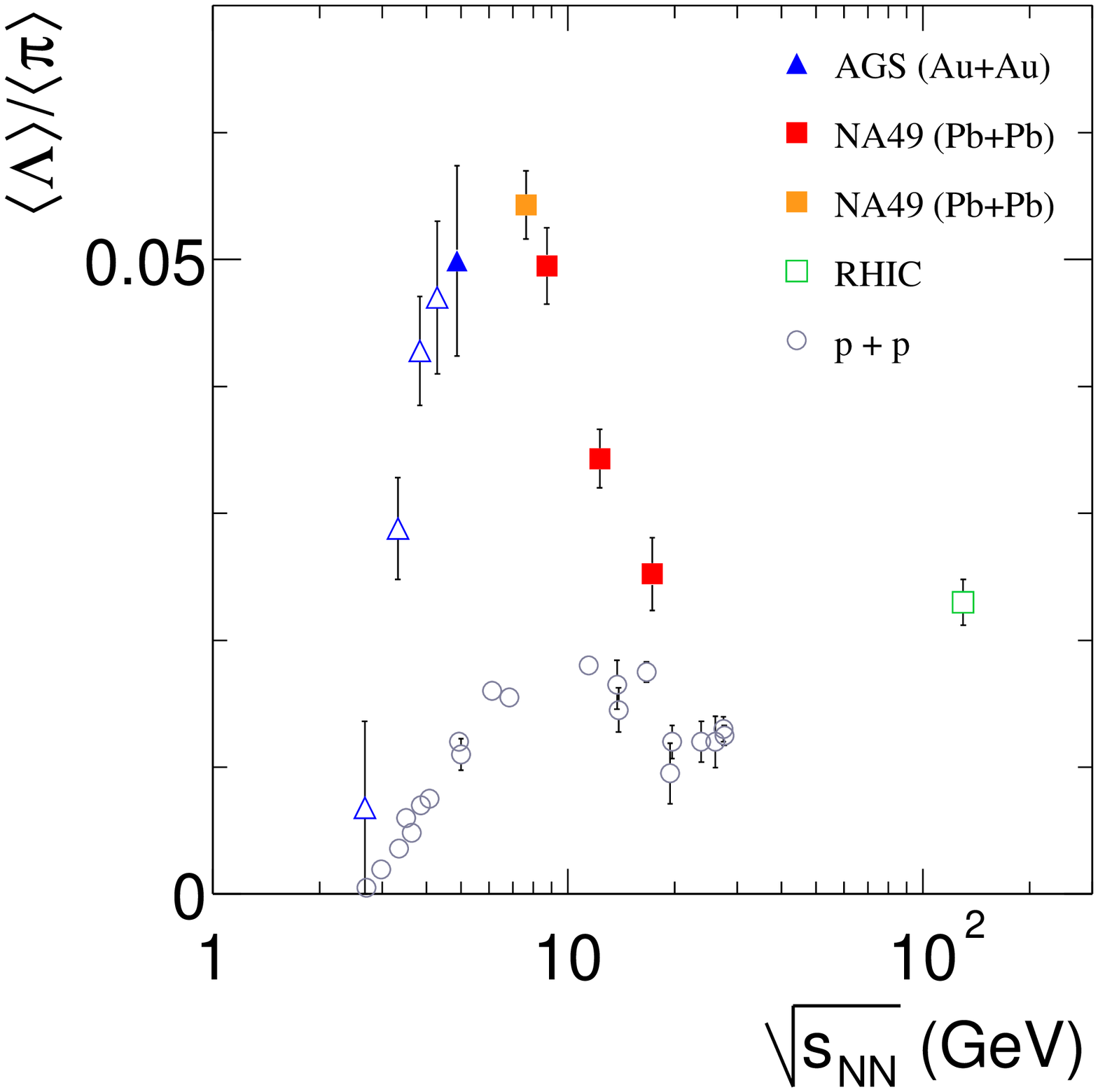,angle=270,width=80mm}
\end{center}
\caption{ Energy dependence of the $<\Lambda>/<\pi>$
($<\pi>=1.5\:(<\pi^->+<\pi^+>)$) ratio for central Pb+Pb (Au+Au)
collisions (upper points) and p+p interactions (lower points).}
\end{figure}

Now we turn to the point of the present study. From a recent
energy scan conducted at the SPS by NA49, studying hadron
multiplicities from $\sqrt{s}$=7 to 17 GeV, a steep maximum was
observed \cite{14} in the $K^+/\pi$ and $\Lambda/\pi$ ratios in
central Pb+Pb collisions, as shown in Figs.1 and 2. As the $K^+$
and $\Lambda$ channels carry most of the total
$<s>+<\overline{s}>$ content, this experimental result indicates a
kind of "singularity" in the strange to non-strange production
ratio, from AGS to RHIC energy. It appears to be unlikely that
state of the art hadronic or partonic quasi-classical microscopic
transport models should exhibit such non-smooth behaviour which is
also absent in $p+p$ collisions. In order to generalize the new
NA49 data, away from consideration of individual channel strange
to non-strange multiplicities, Becattini et al. \cite{3} analyzed
the $\sqrt{s}$ dependence of the Wroblewski-parameter $\lambda_s$
in the grand canonical statistical hadronization model. Their
result is shown in Fig.3 which gives $\lambda_s$ as a function of
the chemical potential $\mu_B$. From top AGS energy (at $\mu_B
\approx$ 550 MeV) to RHIC energy ($\mu_B \leq$ 50 MeV) one
perceives an average $\lambda_s$ of about 0.45 $\pm$ 0.08 whereas
a steep excursion is seen, to $\lambda_s=0.6 \pm 0.1$, at
$\mu_B$=440 MeV. This point corresponds to the steep maxima
observed in Figs.1 and 2, to occur at SPS fixed target energy of
30 GeV/A in central Pb+Pb collisions, corresponding to
$\sqrt{s}$=7.3 GeV. The steep singularity of the NA49 $K^+/\pi$
and $\Lambda/\pi$ ratios at this $\sqrt{s}$ thus reflects in a
maximum of the Wroblewski $\lambda_s$, derived from grand
canonical analysis. For the sake of clarity we state here that the
latter analysis just takes note, in a snapshot at hadronic
freeze-out (to a decoherent hadron/resonance gas) of the general
multiplicity order prevailing at the instant of hadronization. The
observed $\lambda_s$ maximum should, therefore, present a hint
that hadronization at $\sqrt{s} \approx$ 7 GeV should occur under
influences, absent at energies above and below. Moreover, NA49 has
shown recently \cite{15} that the event-by-event fluctuation of
the ratio ($K^+ + K^-)/(\pi^+ + \pi^-$) measured in central Pb+Pb
collisions increases steeply toward $\sqrt{s}$ = 7 GeV whereas it
was formerly found \cite{16} to amount to be below 4\%, at top SPS
energy, $\sqrt{s}$ = 17.3 GeV.

\begin{figure}[ht]
\begin{center}
\epsfig{figure=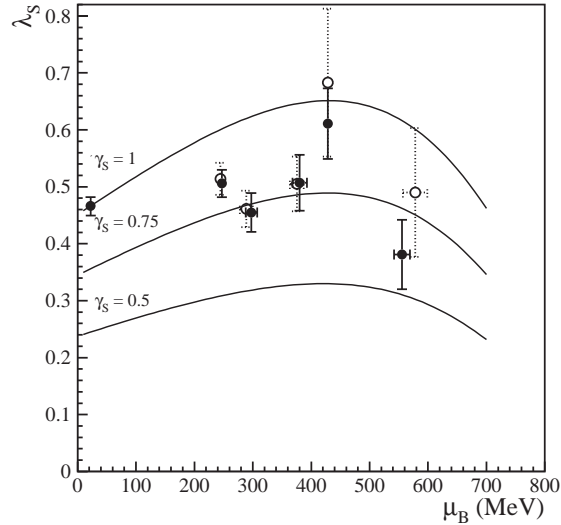,width=8cm} \caption{Dependence
of the $\lambda_s$ parameter on baryochemical potential extracted
from the fits to hadron multiplicities in central Pb+Pb (Au+Au)
collisions at AGS, SPS and RHIC energies. The lines show the
dependence expected for different values of the $\gamma_s$
parameter.}
\end{center}
\end{figure}

We thus propose that the dynamical trajectory of central Pb+Pb
collisions comes close to the critical point of QCD, at or near
$\sqrt{s}$ = 7 GeV. This point has been expected to occur on the
line in the T, $\mu_B$ plane which describes the boundary between
the hadronic and partonic QCD phases \cite{17}. Along that line
the phase transition is expected to be a crossover at
$\mu_B<\mu_B^c$, become second order at $\mu_B = \mu_B^c$, and
first order for $\mu_B > \mu_B^c$. At $\mu_B = \mu_B^c$ and T =
T$_c$ we thus expect phenomena analogous to critical opalescence.
Recent QCD lattice calculations succeeded in an extrapolation to
finite chemical potential \cite{18,19}, thus making a first
prediction for the phase boundary line and, in particular, the
critical point - albeit with considerable uncertainty as was
discussed by Redlich at QM04 \cite{20}. This uncertainty stems,
firstly, from the uncertainty in the extrapolation to finite
$\mu_B$ but, secondly, from the unphysical (high) strange quark
mass employed in these lattice calculations which, at present,
place the critical point somewhere in the interval 500 MeV $ <
\mu_B^c \le$ 700 MeV.  Redlich argued that it should move to
considerably lower $\mu_b$ once the s-mass can be chosen closer to
the physical quark mass \cite{20}. This expectation was
substantiated by recent lattice calculations which show that the
critical point might move downward in $\mu_B$ once more realistic
quark masses are employed \cite{21}. From Fig.3 we see that the
strangeness maximum at $\sqrt{s}$ = 7 GeV corresponds to $\mu_B
\approx$ 440 MeV and thus quite close to the expected $\mu_B^c$
position. Furthermore the energy density at the phase boundary is
estimated by lattice QCD to be rather low \cite{22} ($\epsilon
\le$ 1 GeV/fm$^3$).

Central collisions of heavy nuclei at SPS energy exhibit a general
cycle of initial compression and heating which is followed by a
maximum energy density stage which then turns into expansion and
cooling \cite{23}. The quantities characterizing the overall
system dynamics, such as volume and energy-entropy density etc.
change very rapidly except during the high density stage which
acts analogous to a classical turning point. Also it is only
during this stage that relaxation times can be of comparable
magnitude to the system evolution time scales (like
volume-doubling etc.). One can thus begin only here treating the
system dynamical evolution in terms of energy density, temperature
and chemical potential - thus defining a dynamical trajectory in
the T, $\mu_B$ plane. Only if this turning point coincides closely
with the QCD critical endpoint one could expect to observe
substantial critical phenomena. Now it is well known that the
maximum energy density in central collisions of mass 200 nuclei
amounts (Bjorken estimate) to above 2 GeV/fm$^3$ at top SPS
\cite{24}, and to about 5 GeV/fm$^3$ at RHIC \cite{25} energies,
thus overshooting, by far, the critical QCD energy density. The
system will thus cross the phase coexistence line, upon
re-expansion, whilst already undergoing rapid expansion.
Furthermore, the chemical potential is certainly well below 300
MeV at the time of hadronization.The evolution will thus miss the
critical point at top SPS, and RHIC energies; and at much lower
AGS energies the dynamics falls into the $\mu_B \ge $500 MeV
domain but the energy might not suffice to reach the phase
boundary. In summary we may indeed expect that the dynamical
evolution reaches its energy density plateau phase near the
expected critical point (i.e. at energy density just below or at 1
GeV/fm$^3$, and at $\mu_B$ between 300 and 500 MeV) somewhere in
between maximum AGS and minimum SPS energy.

\begin{figure}[ht]
\begin{center}
\epsfig{figure=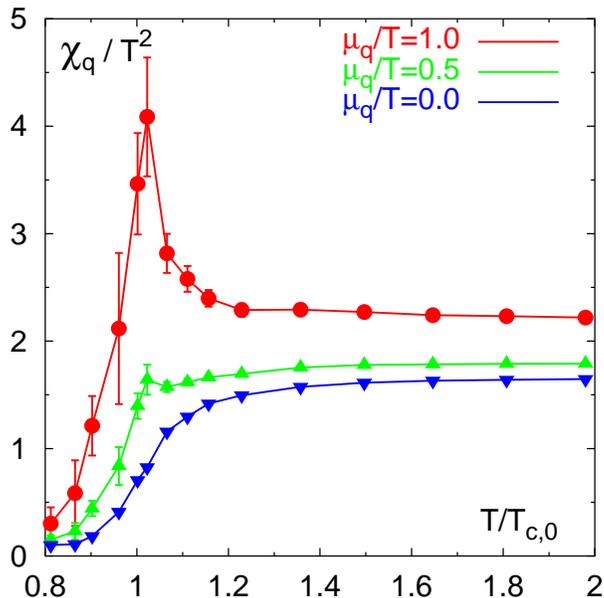,width=8cm} \caption{The quark number
susceptibility calculated within lattice QCD as function of
temperature (relative to transition temperature) for different
values of quark chemical potential.}
\end{center}
\end{figure}

We now turn to the final point of our line of argument by
returning to the lattice results \cite{19,20} at finite $\mu_B$.
They see a steep maximum of the quark number susceptibility
\begin{equation}
\chi_{u,d} \equiv T^2(\frac{d^2}{d(\mu/T)^2} \: \frac{p}{T^4})
\end{equation}

\noindent occuring at T=T$_c$= 150 MeV and $\mu_B = 3 \mu_Q$=3 T =
450 MeV. We reproduce the Bielefeld-Swansea results in Fig.4 which
also shows the calculations for $\mu_B$ = 225 MeV and $\mu_B$ = 0
(essentially corresponding to top SPS and RHIC energies,
respectively). The latter exhibit no susceptibility peak but a
smooth transition from T $<$ T$_c$ to T $ > $ T$_c$. As
$\chi_{u,d}$ can also be written as
\begin{equation}
\chi_q = T^2(\frac{\delta}{\delta(\mu_{u/T})} \:+ \:
\frac{\delta}{\delta(\mu_{d/T})}) \: \frac{n_u + n_d}{T^3}
\end{equation}
we see that the peak in the susceptibility implies a maximum
fluctuation of the quark number densities $n_u$ and $n_d$. We
interpret this result as an indication of critical fluctuation
occuring in the vicinity of the critical endpoint implicitly
present in this calculation. Directly at $\mu^c$ the
susceptibility would diverge. The critical point in this
calculation must thus be near $\mu_B$ = 450 MeV and T = 150 MeV.
This, in turn, is very close to the parameters of the grand
canonical model at the strangeness maximum, $\sqrt{s}$ = 7 GeV
(Fig.3).

As to the relation between the susceptibility maximum of lattice
QCD and the strangeness maximum observed by NA49 (at which the
Wroblewski parameter $\lambda_s$ exhibits an anomaly), Gavai and
Gupta \cite{26} have suggested the relationship
\begin{equation}
\lambda_s = \frac{2 \chi_s}{\chi_u+\chi_d}
\end{equation}
which appears to offer a direct link. In fact they obtain
$\lambda_s$ = 0.48 from a lattice calculation at zero $\mu_b$:
closely coinciding with the value observed at top SPS and RHIC
energy (Fig.3). Unfortunately, though, their result refers to
$\mu_b$ = 0, and the Bielefeld-Swansea calculations at finite
$\mu_b$ \cite{19} are in two-flavour QCD only. A prediction for
$\chi_s$ at $\mu_B \approx $450 MeV, or, more generally, a full
three-flavour lattice treatment of the vicinity of the critical
point is required to finally assess the line of argument of the
present note. If proven correct we would encounter, here, the
first {\bf direct} reflection of QCD in nucleus-nucleus collision
data.

The author acknowledges stimulative discussions with F. Becattini,
R. V. Gavai, F. Karsch and K. Redlich.


\begin{thebibliography}{99}
\bibitem{1} F. Becattini, M. Gazdzicki and J. Sollfrank, Nucl.
Phys. A638 (1998) 403;\\
J. Cleymans and K. Redlich, Phys. Rev. Lett. 81 (1998) 5284.
\bibitem{2} P. Braun-Munzinger, I. Heppe and J. Stachel, Phys.
Lett. B465 (1999) 15;\\
P. Braun-Munzinger, D. Magestro, K. Redlich and J. Stachel, Phys.
Lett B518 (2001) 41;\\
P. Braun-Munzinger, K. Redlich and J. Stachel, nucl-th/0304013;\\
P. Braun-Munzinger, J. Cleymans, H. Oeschler and K. Redlich, Nucl.
Phys. A697 (2002) 902.
\bibitem{3} F. Becattini, M. Gazdzicki, A. Keraenen, J. Manninen
and R. Stock, Phys. Rev. C69 (2004) 024905.
\bibitem{4} R. Hagedorn, Nucl. Phys. B24 (1979) 93;\\
J. Ellis and K. Geiger, Phys. Rev. D54 (1996) 1967.
\bibitem{5} R. Stock, Phys. Lett B456 (1999) 277;\\
R. Stock, hep-ph/0312039.
\bibitem{6} R. Hagedorn, Nuovo Ciemento 35 (1965) 395.
\bibitem{7} F. Becattini, Z. Phys. C69 (1996) 485.
\bibitem{8} F. Becattini and L. Ferroni, hep-ph/0307061.
\bibitem{9} J. Cleymans et al., Phys. Rev. C56 (1997) 2747;\\
A. Tounsi and K. Redlich, J. of Physics G28 (2002) 2095.
\bibitem{10} S. V. Afanasiev et al., NA49 Coll., Phys. Rev. C66
(2002) 054902;\\
A. Mischke et al., NA49 Coll., Nuc. Phys. A715 (2003) 453.
\bibitem{11} L. Ahle et al., E802 Coll., Phys. Rev. C60 (1999)
064901.
\bibitem{12} F. Antinori et al., NA57 Coll., Nucl. Phys. A698
(2002) 118;\\
F. Antinori, talk at QM04.
\bibitem{13} A. K. Wroblewski, Acta Phys. Pol. B16 (1985) 379.
\bibitem{14} V. Friese et al., NA49 Coll., nucl-ex/0305017;\\
M. Gazdzicki, talk at QM04.
\bibitem{15} Ch. Roland et al., NA49 Coll., talk at QM04.

\bibitem{16} S. V. Afanasiev et al., NA49 Coll., Phys. Rev. Lett.
86 (2001) 1965.
\bibitem{17} M. G. Alford, K. Rajagopal and F. Wilczek, Phy. Lett
B442 (1998) 247;\\
Y. H. Atta and T. Ikeda, Phys. Rev. D67 (2003)
014028.
\bibitem{18} Z. Fodor and S. D. Katz JHEP 0203 (2002) 14.
\bibitem{19} R. V. Gavai and S. Gupta, hep-lat/0303013;\\
C. R. Allton et al., hep-lat/0305007.


\bibitem{20} K. Redlich, talk at QM04.
\bibitem{21} Z. Fodor and S. D. Katz, hep-lat/0402006;\\
F. Karsch et al., Bielefeld-Swansea lattice Collaboration,
hep-lat/0309116.
\bibitem{22} F. Karsch, Nucl. Phys. A698 (2002) 199.
\bibitem{23} R. Stock, Phys. Reports, 135 (1986] 259.
\bibitem{24} T. Alber et al., NA49 Coll., Phys. Rev. Lett 75
(1995) 3814.
\bibitem{25} G. Roland, talk at QM04.
\bibitem{26} R. V. Gavai and S. Gupta, Phys. Rev. D (2002);\\
R. V. Gavai, talk at QM04;\\
R. V. Gavai and S. Gupta, hep-lat/0303013.
\end{thebibliography}
\end{document}